\documentclass[twocolumn]{aastex62}

\hypersetup{linkcolor=red,citecolor=green,filecolor=cyan,urlcolor=magenta}

\graphicspath{{./}{figures/}}

\received{xxx xx, 2019}
\revised{xxx xx, 2019}
\accepted{xxx xx, 2019}

\submitjournal{ApJL}

\shorttitle{Non-lognormal Markov model}
\shortauthors{Mocz \& Burkhart}

\begin{document}

%
%
%
%
%
%
%
%

\title{A Markov model for non-lognormal density distributions in compressive isothermal turbulence}

\correspondingauthor{Philip Mocz}
\email{pmocz@astro.princeton.edu}

\author[0000-0001-6631-2566]{Philip Mocz$^{*}$}
\affil{Department of Astrophysical Sciences, Princeton University, 4 Ivy Lane, Princeton, NJ, 08544, USA}
\altaffiliation{Einstein Fellow}

\author[0000-0001-5817-5944]{Blakesley Burkhart}
\affil{Center for Computational Astrophysics, Flatiron Institute, 162 Fifth Avenue, New York, NY 10010, USA}
\affil{Department of Physics and Astronomy, Rutgers University,  136 Frelinghuysen Rd, Piscataway, NJ 08854, USA}

\begin{abstract}

Compressive isothermal turbulence is known to have a near lognormal density probability distribution function (PDF) with a width that scales with the sonic Mach number and nature of the turbulent driving (solenoidal vs compressive). However, the physical processes that mold the extreme high and low density structures in a turbulent medium can be different, with the densest structures being composed of strong shocks that evolve on shorter timescales than the low density fluid. The density PDF in a turbulent medium exhibits deviations from lognormal due to shocks, that increases with the sonic Mach number, which is often ignored in analytic models for turbulence and star formation. We develop a simple model for turbulence by treating it as a continuous Markov process, which explains both the density PDF and the transient timescales of structures as a function of density, using a framework developed in \cite{2018ApJ...865L..14S}. Our analytic model depends on only a single parameter, the effective compressive sonic Mach number, and successfully describes the non-lognormal behavior seen in both 1D and 3D simulations of supersonic and subsonic compressive isothermal turbulence. The model quantifies the non-lognormal distribution of density structures in turbulent environments, and has application to star forming molecular clouds and star formation efficiencies.

\end{abstract}


\section{Introduction} \label{sec:intro}

The interstellar medium (ISM) is a supersonic, highly turbulent, magnetized medium \citep{1981MNRAS.194..809L,1987ARA&A..25...23S}. 
The turbulent density distribution plays a crucial role in the process of star formation, including setting the initial mass function, star formation rates and star formation efficiency.  
There is a well established numerical and analytic result that the density distribution of a supersonic, isothermal medium is approximately lognormal, with a variance that grows with the sonic Mach number $\mathcal{M}$
\citep{1998PhRvE..58.4501P,2011ApJ...730...40P,2011ApJ...727L..21P,2012MNRAS.423.2680M, 2012ApJ...755L..19B}.
Past a critical density threshold, the high density tail of the distribution may experience collapse under self-gravity, and form stars. It is thus important to understand the distribution and timescales of high density structures from a theoretical perspective, to understand the initial conditions of star formation and predict star formation efficiencies.

Despite the wide usage of the lognormal model \citep{1994ApJ...423..681V,2008ApJ...688L..79F,2011ApJ...727L..21P,2013MNRAS.430.1880H}, supersonic turbulence is known to also exhibit deviations from a lognormal density PDF deviations from a lognormal density PDF due to the presence of strong shocks and intermittency \citep{2007ApJ...658..423K,2013MNRAS.436.1245F,2017MNRAS.471.3753S,2019arXiv190500923P}. The deviation becomes stronger at higher Mach numbers (i.e., for stronger density compressions and larger degree of density contrast).
The non-lognormal density PDF can be described well with a simple spatial model that considers the density to be arranged as a collection of strong shocks of width $\mathcal{M}^{-2}$ \citep{2017MNRAS.471.3753S}.

Recently, \cite{2018ApJ...854...88R} put forth the theoretical picture (confirmed by simulations) that a supersonic, turbulent, isothermal medium is a network of quasi-1D strong shock sheet fragments with exponential post-shock profiles\footnote{see also \citealt{2018MNRAS.480.3916M} for the magnetohydrodynamic version of the picture}. These strong shocks make up most of the mass fraction (and high density tail of the density PDF), but fill just a tiny fraction of the volume. The high density shocks are transient structures, that evolve and dissipate on faster timescales than the cloud crossing time (but see also \citealt{2008ApJ...679..537F}). 

A complimentary picture to describing the density PDF as a spatial composition as was done in \cite{2018ApJ...854...88R}, is to consider the statistical time-evolution of the densities of fluid elements.
The volume elements that make up the density PDF, for example may be considered a Markov process
(similarly, the velocity field in a turbulent Navier-Stokes velocity field can be described by a Markov process \citealt{novikov1989two,pedrizzetti1994markov,pedrizzetti1999quadratic,renner2001experimental}). Indeed, a lognormal distribution is suggestive of an Ornstein-Uhlenbeck stochastic process, i.e., a mean-reverting random walk \citep{1930PhRv...36..823U}. The density PDF in supersonic turbulence has been modelled as such in
\cite{2016MNRAS.460.4483K}. 
More recently, \cite{2018ApJ...865L..14S} considered a Markov model for explaining not only the turbulent PDF, but also the rate of change of the log density ($s$) over a finite time interval, given as a function of final value of $s$. Here we adopt a similar approach to study the non-lognormality of the density PDF. The important differences between our model and theirs are: (1) we seek to describe the non-lognormality of the density PDF whereas \cite{2018ApJ...865L..14S} was designed to predict near lognormal distribution, (2) our model is fully analytic, (3) we describe the behavior in the Eulerian frame as opposed to Lagrangian, and (4) we ensure exact mass conservation with proper renormalization.

Thus our focus is to construct a simple, predictive Markov model to describe the non-lognormal distribution of density in compressive turbulence. We do so by applying simple assumptions for the timescales of evolving structures as a function of density. We then measure the timescales in 1D simulations of isothermal `turbulence' -- such simulations are entirely shock-dominated, as there is no true turbulence in 1D. Thus, the simple set up we work with here is akin to Burgers turbulence. These 1D simulations allow us to fit the free parameters of our model and study the dependence on the sonic Mach number. We then compare our model to 3D simulations, which exhibit an interplay between strong shocks and a turbulent cascade. 

We lay out our notation and a brief overview of the isothermal fluid equations and the lognormal model in \S~\ref{sec:eqn} and \ref{sec:pdf}.
We describe out analytic Markov model for transient shocked-structures turbulence in \S~\ref{sec:model}, as well as fitting the model parameters with 1D simulations.
We compare our model to previous models in the literature for the PDF, and to results from 3D supersonic isothermal turbulence simulations in \S~\ref{sec:disc}. We summarize our main findings in \S~\ref{sec:conc}.

\section{Isothermal fluid equations}\label{sec:eqn}

The equations for an isothermal, inviscid fluid can be written compactly as follows.
Let $\overline{\rho}$ be the average density in the medium, 
and use units where the sound-speed is $c=1$. The equation of state of such a fluid is $P=\rho c^2 = \rho$.
We define the log-density as: $s\equiv \log(\rho/\overline{\rho})$.
Then, the isothermal fluid equations are:
\begin{equation}
    D_ts=-\nabla\cdot\mathbf{v}
    \label{eqn:compress}
\end{equation}
\begin{equation}
    D_t\mathbf{v}=-\nabla s
    \label{eqn:momentum}
\end{equation}
where $D_t = \partial_t + \mathbf{v}\cdot \nabla$ is the convective derivative.

In the Lagrangian frame, density changes are entirely due to the divergence of the velocity field (compressibility of the fluid).

In the Eulerian frame, the corresponding equations are:
\begin{equation}
    \partial_t s = - \mathbf{v}\cdot \nabla s -\nabla\cdot\mathbf{v}
\end{equation}
\begin{equation}
    \partial_t \mathbf{v} = - \mathbf{v}\cdot \nabla\mathbf{v} -\nabla s
\end{equation}
Changes in density at a fixed location are due to a combination of advection and compression.

\section{Log-density PDF}\label{sec:pdf}

Our object of study is the density PDF.
Consider the \textit{volume-weighted} PDF of the log-density, $P_V(s)$, which is the probability of a given unit volume $V$ having an average density $\rho=\exp(s)$, per unit $s$.

The PDF must obey 2 constraints, the conservation of probability,
\begin{equation}
\int_{-\infty}^\infty P_V(s)\,ds = 1,
\label{eqn:probconstraint}
\end{equation}
and the conservation of mass,
\begin{equation}
\int_{-\infty}^\infty \rho P_V(s)\,ds = \overline{\rho} = 1.
\label{eqn:massconstraint}
\end{equation}

We can also define the related \textit{mass-weighted} density PDF, as $P_M(s) = \rho P_V(s)$.

A common model for the density PDF is the lognormal:
\begin{equation}
    P_V^{\rm LN}(s) = \frac{1}{\sqrt{2\pi\sigma^2}}
    \exp\left[-\frac{(s-s_0)^2}{2\sigma^2}\right]
    \label{eqn:LN}
\end{equation}
where $s_0=-\sigma^2/2$ is the center of the normal distribution, and $\sigma^2$ is its variance.
The variance has been empirically determined from numerical simulations of driven isothermal turbulence to be a function of the sonic Mach number and the turbulent driving (b):
\begin{equation}
    \sigma^2 = \log(1+b^2\mathcal{M}^2)
\end{equation}
where $b\simeq1$ for compressive driving and 
$b\simeq1/3$ for solenoidal \citep{2008ApJ...688L..79F,2011ApJ...727L..21P}.
If the volume-averaged PDF is lognormal, the mass-weighted PDF is too, just with shifted mean: $s_0\to\sigma^2/2$.

A well-known motivation for the origin of a lognormal PDF comes from considering the statistics of the divergences in the velocity field in the right-hand-side of Eqn.~\ref{eqn:compress} responsible for evolving $s$ and applying the Central Limit Theorem \citep[e.g.][]{1994ApJ...423..681V}. If a Lagrangian fluid element of log-density $s$ encounters instances of diverging velocity fields in the fluid which behave like independent random variables, then by the Central Limit Theorem, the mass-weighted distribution of the log-density $s$ is normal, and thus so is the volume-weighted distribution. Any deviation from the assumption of independent random variables will lead to deviations in the lognormal distribution.

\begin{figure}[ht!]
\plotone{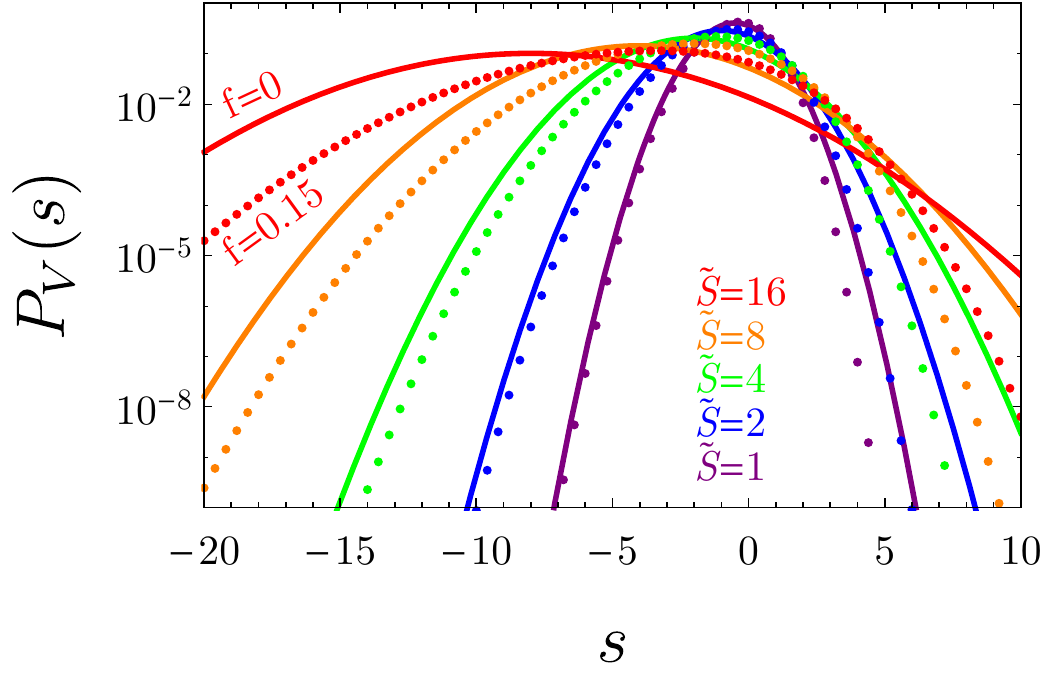}
\plotone{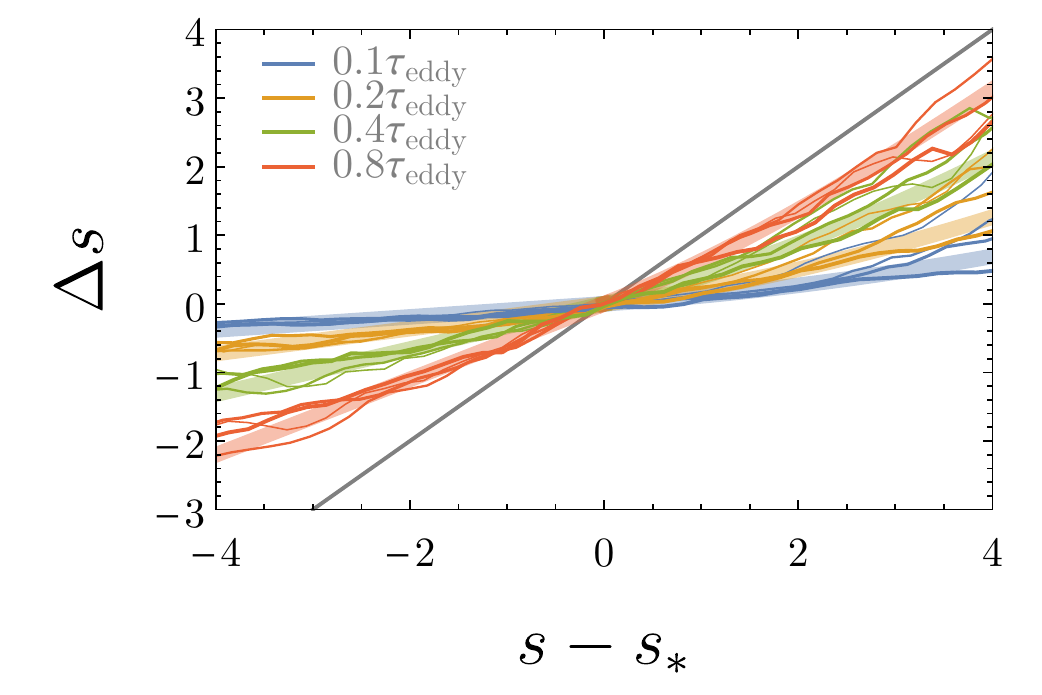}
\caption{\textit{Top panel}: A plot of our model for the density PDF (Eqn.~\ref{eqn:main}). $f=0$ (solid) is a simple Gaussian, 
while $f>0$ introduces negative skewness (shown in dotted lines is $f=0.15$). 
\textit{Bottom panel}:
Measure of the density-dependent transient timescales in 1D isothermal shock simulations used to tune our model. $\Delta s$ reverts back to the straight gray line after $\sim \tau_{\rm eddy}$, meaning that the structure at a given location is now uncorrelated with what used be there. 
Shown is our model fit (very thick, transparent line) and simulation data ($\mathcal{M}=2$ (thin), $4$, $8$, $16$ (thick)) at $4$ different times.
High density structures tend to revert back to mean distribution (gray line) faster than structure at low densities. 
\label{fig:model}}
\end{figure}

\section{A Markov model for non-lognormal density distribution in compressible turbulence}  \label{sec:model}

We model the volume-weighted density PDF in the Eulerian frame as a Markov process:
\begin{equation}
s(t+dt) = s(t) + A(s)dt + \mathcal{N}(0,1)\sqrt{D(s)dt}
\end{equation}
where
\begin{equation}
    A(s) = \frac{s_*-s}{\tau_A(s)}
\end{equation}
is a drift term that returns $s$ towards the mean $s_*$ of the distribution on a timescale of $\tau_A(s)$, and 
\begin{equation}
    D(s) = \frac{2\sigma_s^2}{\tau_D(s)}
\end{equation}
is is a diffusion term, with timescale $\tau_D(s)$
and $\mathcal{N}(0,1)$ is a random number drawn from a normal distribution of mean zero and variance one.

In the case that $\tau_A(s)=\tau_D(s)={\rm const}$, 
the Markov process is the well-known Ornstein-Uhlenbeck process (mean-reverting random walk), 
and results in a lognormal distribution for $s$ (Eqn.~\ref{eqn:LN}). Having constant drift and diffusion timescales is the unique solution that predicts a lognormal distribution, although modifications to the timescales can yield distributions that are very close to lognormal \citep{2018ApJ...865L..14S}. 

Despite the turbulent density PDF being near lognormal, the types of structures at high and low densities are quite different. High density gas is composed of strong shocks arranged in quasi-1D sheet fragments that can interact. They have a large mass fraction but small volume-filling fraction.
Low density regions are created by the interaction of 3-D expansion waves, structured as large volume-filling irregular voids. The high density part of the PDF establishes itself first very quickly when the medium is turbulently stirred, in just a fraction of the dynamical time \citep{1999intu.conf..218N}. These shocks are known to operate on faster timescales than the rest of the fluid \citep{2018ApJ...865L..14S,2018ApJ...854...88R}.

Thus we consider a drift timescale that is constant at low densities $s<s_*$, but decreases at high densities $s>s_*$:
\begin{equation}
    \tau_A(s) = \frac{\tau_{A,0}}{1+{\rm H}(s-s_*)3f/2}
\end{equation}
where ${\rm H}(s)$ is the Heaviside step function, $\tau_{A,0}$ is a characteristic timescale, and $f\geq 0$ is a free-parameter setting how strongly the timescale decreases at high densities.
This is a phenomenological model meant to capture the fact that strong shocks can form, expand, and revert back to the mean on faster timescales than the typical dynamical time of the system, but it is rather quite general to describe a two-state behaviour which ends up yielding tractable analytic solutions.
We also consider a constant dispersion term:
\begin{equation}
    D(s) = \frac{2\sigma_s^2}{\tau_{D,0}}
\end{equation}
where $\sigma_s$ is the size of the dispersion and $\tau_{D,0}$ is the associated dispersion timescale.

This Markov process has a steady-state analytic solution for the density PDF, which can be obtained from solving the time-steady Fokker-Planck equation:
\begin{equation}
    0=-\frac{\partial}{\partial s}\left[A(s)P(s)\right] + \frac{\partial^2}{\partial s^2}\left[\frac{D(s)}{2}P(s)\right]
\end{equation}

The solution is:
\begin{equation}
    P_V(s) \propto \exp\left[
    -\frac{(s-s_*)^2\left(1+f(s-s_*){\rm H}(s-s_*)\right)}{2\tilde{S}}.
    \right]
    \label{eqn:main}
\end{equation}
where we have defined
\begin{equation}
    \tilde{S}\equiv \sigma_s^2\frac{\tau_{A,0}}{\tau_{D,0}}
\end{equation}
The solution depends only on the drift-to-dispersion timescale ratio, $\tau_{A,0}/\tau_{D,0}$, not their absolute values.

The normalization of Eqn.~\ref{eqn:main} is computed from the probability constraint (Eqn.~\ref{eqn:probconstraint}), and $s_*$ from the mass constraint (Eqn.~\ref{eqn:massconstraint}). The analytic expressions are a bit cumbersome but it is straightforward to compute these values numerically.
Note that the lognormal distribution (Eqn.~\ref{eqn:LN}) is recovered when $f=0$ and $\tau_{A,0}=\tau_{D,0}$.

The model, at this stage, has two free parameters that need to be fit to data: $f$, a knob for the degree of non-lognormality, and $\tilde{S}$ which sets the width of the distribution.
In the top panel of Fig.~\ref{fig:model} we plot the shapes of the predicted PDFs for $f=0.15$ and compare it to $f=0$ (lognormal), for various widths $\tilde{S}$. 
The $f>0$ parameter introduces a degree of negative skewness. The deviation from lognormal grows stronger as the PDF is made wider.
We will fit the free parameters to 1D `turbulence' simulations, described in sections~\ref{sec:sims} and \ref{sec:fit}.


\begin{figure}[ht!]
\plotone{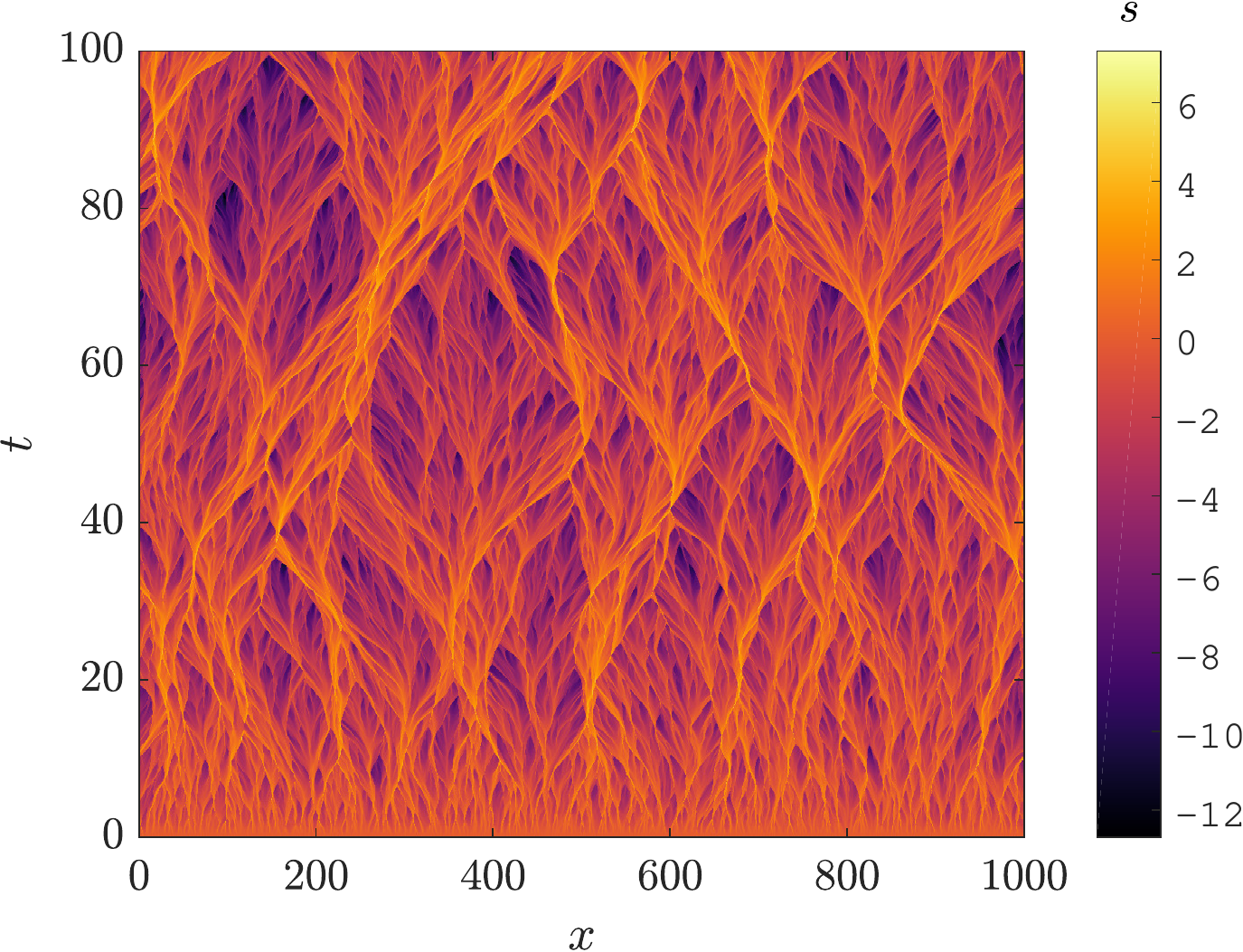}
\plotone{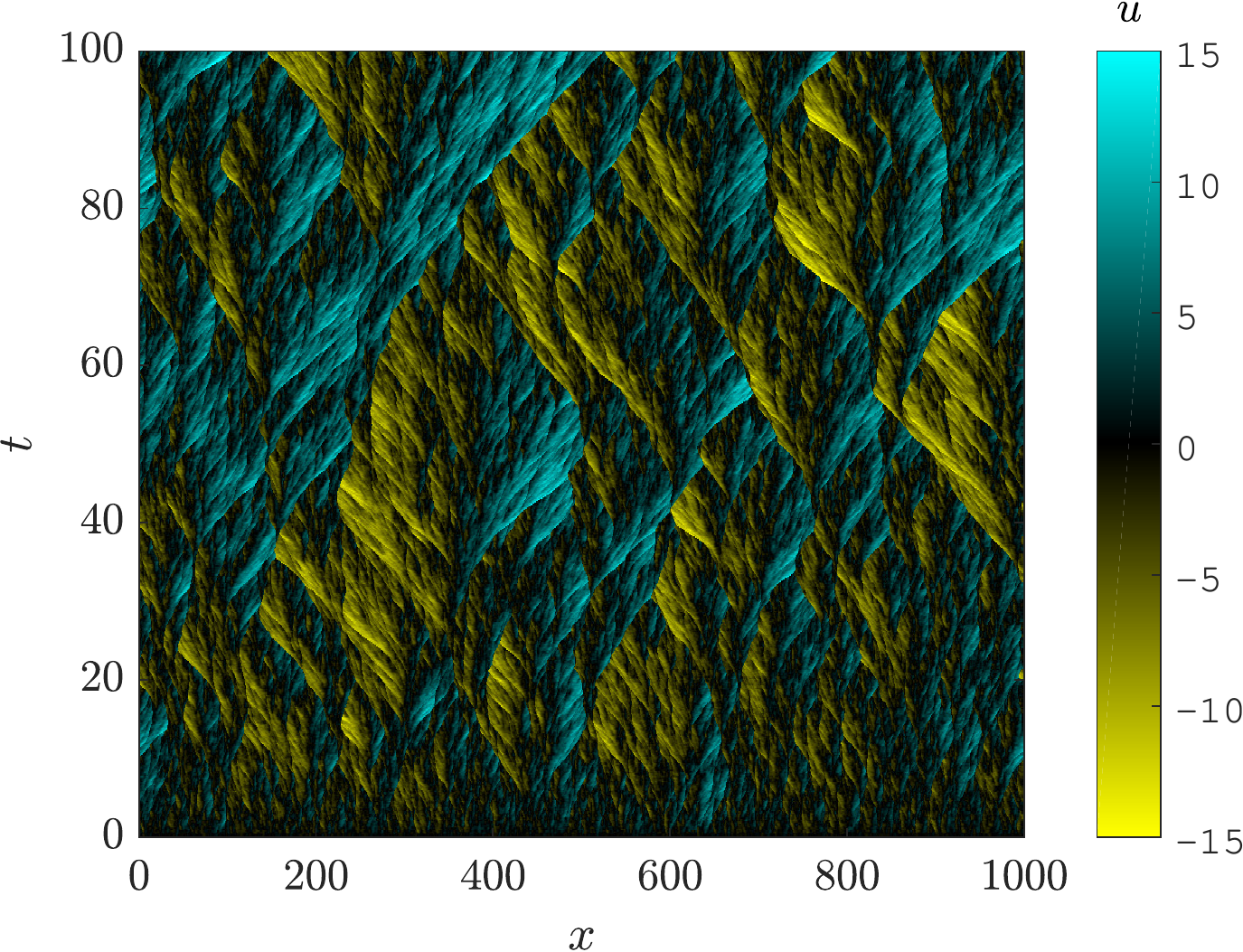}
\caption{Space-time diagram for 1D isothermal supersonic $\mathcal{M}=4$ `turbulence' (\textit{top}: density, \textit{bottom}: velocity). The system quickly establishes a network of self-similar interacting and expanding shocks.
}
 \label{fig:spacetime}
\end{figure}

\subsection{1D isothermal `turbulence' simulations}\label{sec:sims}

We carry out 1D isothermal `turbulence' simulations to which we will tune the Markov model. 
This section is dedicated to the numerical implementation details.
We solve Eqns.~\ref{eqn:compress} and \ref{eqn:momentum} using a 2nd order finite volume (Godunov) method, following \cite{2010MNRAS.401..791S} on a Cartesian grid.
We add a random driving term to the right-hand-side of Eqn.~\ref{eqn:momentum} to drive the turbulence.

Shocks are driven on a scale of $L_{\rm stir}=1$. The box size is much larger, $L_{\rm box}=1000$, to capture the statistics of shocks. The resolution is $\Delta x = 0.0625$ to resolve shocks. 

The shocks are driven by a random driving term, which has to be compressive in 1D. The details of the driving do not matter significantly in 1D for the statistical properties of the shocks, unlike in 3D where they can be a mix of compressive and solenoidal. 
We drive the turbulence by adding $5\times L_{\rm box}$ Gaussian pulses and uniform random locations $x_{\rm rand,L_{\rm box}}$ in the box, in time intervals of $\Delta t_{\rm kick}=0.1$:
\begin{equation}
    v_{\rm source} = \sum_{{\rm shock}=1}^{5\times L_{\rm box}} \mathcal{A}\times \mathcal{N}(0,1) \exp\left(-x_{\rm rand,L_{\rm box}}/\left(2L_{\rm stir}^2\right)\right)
\end{equation}
where $\mathcal{A}$ is the amplitude of the kicks, which sets the resulting turbulent Mach number. We simulate turbulence with mach numbers $\mathcal{M}=2,4,8,16$.

We find the driving establishes a quasi-steady state solution (approximately constant Mach number) where the energy injected is dissipated by the shocks.
The result is `burgulence': a network of 1D shocks (saw-tooth shaped as a function of $s$), with a velocity power spectrum of $k^{-2}$ as in Burgers' turbulence, as opposed to the $k^{-5/3}$ Kolmogorov result from 3D incompressible turbulence. 
Fig.~\ref{fig:spacetime} shows a space-time diagram of the full solution from $t=0$ to $t=100$.

\subsubsection{Fitting the Markov model timescales from 1D simulations}\label{sec:fit}

Our Markov model is best fit with $f=0.15$ and $\tilde{S} = 1.4\mathcal{M}$ with the 1D simulations (bottom panel Fig.~\ref{fig:model}).

We have carried out a suite of 1D driven isothermal `turbulence' simulations, described in \S~\ref{sec:sims}, with sonic Mach numbers of $\mathcal{M}=2,4,8,16$. The simulations have a characteristic timescale $\tau_{\rm eddy}=\frac{L_{\rm stir}}{\mathcal{M}c}$ associated with the stirring length scale.

We use the simulations to fit $\tau_{A,0}$ and $f$, and found $\tau_{A,0}=\tau_{\rm eddy}$ and $f=0.15$.
The fit was performed as follows, using an approach similar to that in \cite{2018ApJ...865L..14S} (that work instead treated Lagrangian tracer particles rather than volume elements). We considered the average change $\Delta s$ of an Eulerian volume element over a time interval, given a final value of $s_2$. The average change of our Markov process should evolve as
\begin{equation}
    \Delta s(s_2,\Delta t)=
    \left(1-\exp\left[-\Delta t/\tau_A(s_2)\right]\right)\left(s_2-s_*\right)
    \label{eqn:ds}
\end{equation}
The relation approaches $s_2-s_*$ at long times; that is, after an eddy time the fluid of a given density $s_2$ is expected to lose memory of its past and on average revert back to the mean log-density. 
We measured $\Delta s$ as a function of time and density in our 1D simulations of varying Mach number, and fit it with Eqn.~\ref{eqn:ds}, see Fig.~\ref{fig:model} (bottom panel).
This idea of measuring $\Delta s$ comes from \cite{2018ApJ...865L..14S}, where in that work it was measured in a Lagrangian frame.

Then, we fit the term controlling the diffusion, $\tilde{S}$, to match the resulting PDFs of the simulations. We find $\tilde{S} = 1.4\mathcal{M}$ for our 1D simulations.

In 1D, our model reduces to a single parameter model (the sonic Mach number $\mathcal{M}$).
We shall see in \S~\ref{sec:apply} that in 3D our model still works remarkably well, if $\mathcal{M}$ is replaced with the effective compressive Mach number.

\begin{figure*}[ht!]
\plotone{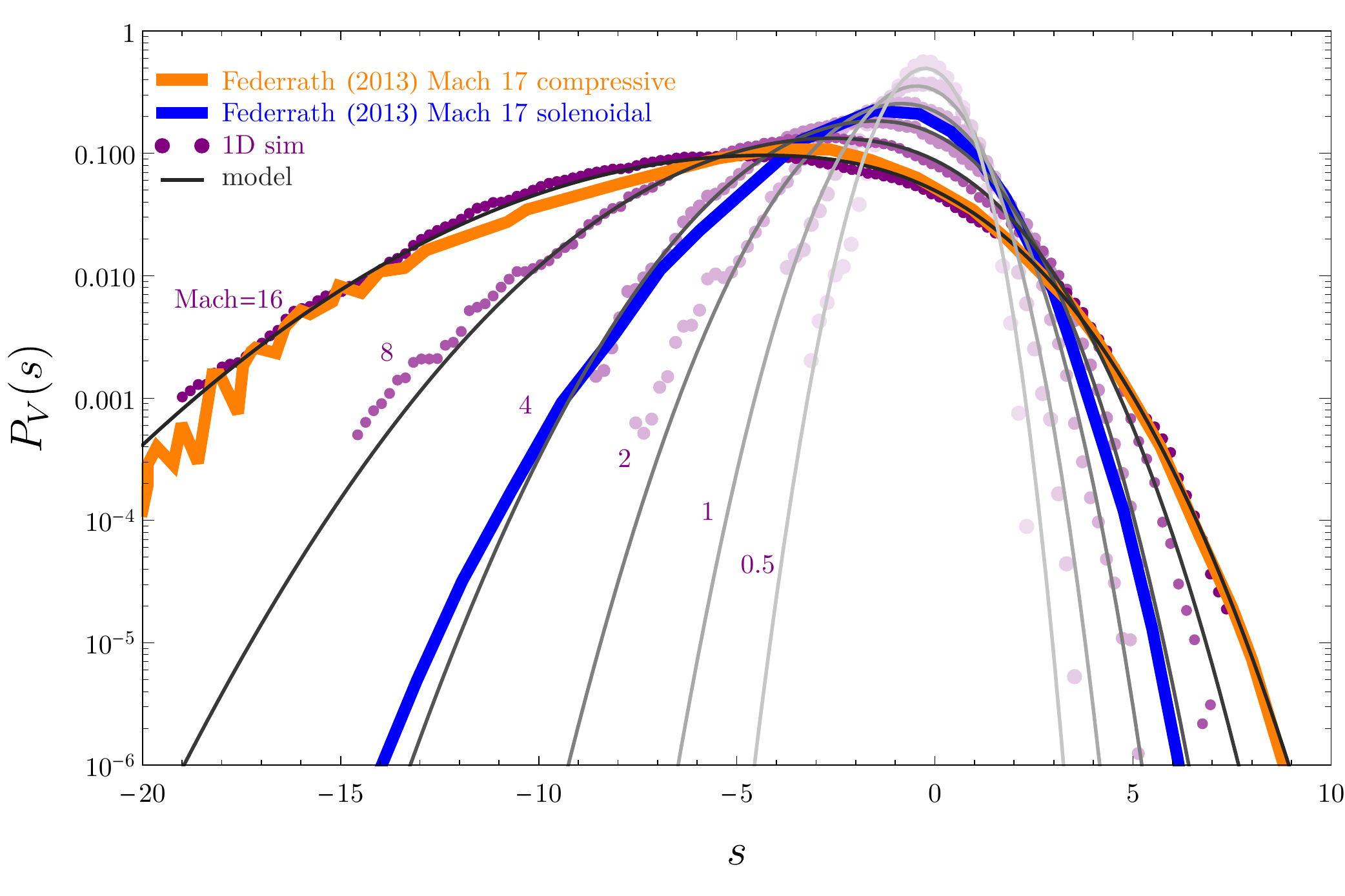}
\caption{Volume-average density PDFs of our Markov model and simulations. We compare our Markov model prediction for the PDF (gray lines) to 1D `turbulence' shock simulations (purple dots) with Mach numbers $\mathcal{M}=0.5,1,2,4,8,16$. The distributions are close to lognormal, but strong shocks lead to negative skewness that increases with Mach number. Our models also do a good job at explaining the PDFs of 3D turbulent simulations. Shown here are two $\mathcal{M}\sim 17$ high-resolution ($4096^3$) simulations from \cite{2013MNRAS.436.1245F}, one driven solenoidally, the other compressively. In the compressive case, the PDF looks similar to a 1D simulation with a similar Mach number. In the solenoidal case, the PDF looks like that of a 1D simulation with the Mach number reduced by a factor of $4$.
\label{fig:fit}}
\end{figure*}
\section{Results}\label{sec:disc}

\subsection{Applying the Markov model to turbulent simulations} \label{sec:apply}

We investigate how well our Markov model can predict the non-lognormal density PDF of 1D and 3D turbulent simulations. 
In Fig.~\ref{fig:fit}, 
we verify that our model, calibrated to the 1D simulations, also predicts the resulting density PDF of the 3D simulations accurately. 

We plot the model and 1D simulation data for Mach numbers $\mathcal{M}=2,4,8,16$. As the Mach number increases, the density distribution becomes wider, has a more negative mean, and a more negative skewness (due to strong shocks).  
In addition, we have carried out 1D simulations with Mach numbers $\mathcal{M}=0.5,1$ to show that our model is also applicable to the compressive subsonic and transonic case, where our model reduces to a near lognormal.

We also show the PDFs of very high resolution ($4096^3$) 3D isothermal turbulent simulations with $\mathcal{M}\sim 17$ from \cite{2013MNRAS.436.1245F}.
The tails of the PDF can be tricky to measure due poor statistics at lower resolution, which is why we rely on this high resolution run here for our comparison\footnote{see \cite{2013MNRAS.436.1245F} for a discussion of how finite resolution biases the PDF}. There are two different turbulent driving modes simulated: fully compressive, and fully solenoidal. 
The compressive driving exhibits a much wider and more non-lognormal density distribution than does the solenoidal driving.
In the compressive case, the PDF looks similar to a 1D simulation with the a similar Mach number ($\mathcal{M}\sim 16$). In the solenoidal case, the PDF looks like that of a 1D simulation with the Mach number reduced by a factor of $4$. These are rather remarkable results.
That is, the 3D simulations can be modeled with our Markov model (with $f=0.15$ and $\tilde{S}=1.4\mathcal{M}$ as in the 1D case) if the Mach number $\mathcal{M}$ is replaced with an effective compressive Mach number describing the system, which we point out is a function of the sonic Mach number and the details of the turbulence driving.
In our  1D modeling, there are only compressive (shock) motions. In 3D, even if the large-scale driving is compressive, shock collisions will generate vorticity.
If the driving is solenoidal, the vorticity in the fluid is larger (by a factor of $\sim 2$; \citealt{2013MNRAS.436.1245F}), because solenoidal motions are directly injected. But in the turbulent cascade, some of those motions are converted into compressive motions and shocks form.
The amount of compressive motion will be smaller in solenoidally driven turbulence, and thus the strength of shocks will be reduced.

While we have not compared the timescale $\Delta s$ (which is a function of $\tau_A$) in the 3D simulations of \cite{2013MNRAS.436.1245F} directly to our model, 
we point out that we do not expect it to be different from the 1D simulations, up to constant factor, 
otherwise the resulting density PDF would also change significantly unless $\tau_D$ somehow conspires to give the same shape.

\begin{figure}[ht!]
\plotone{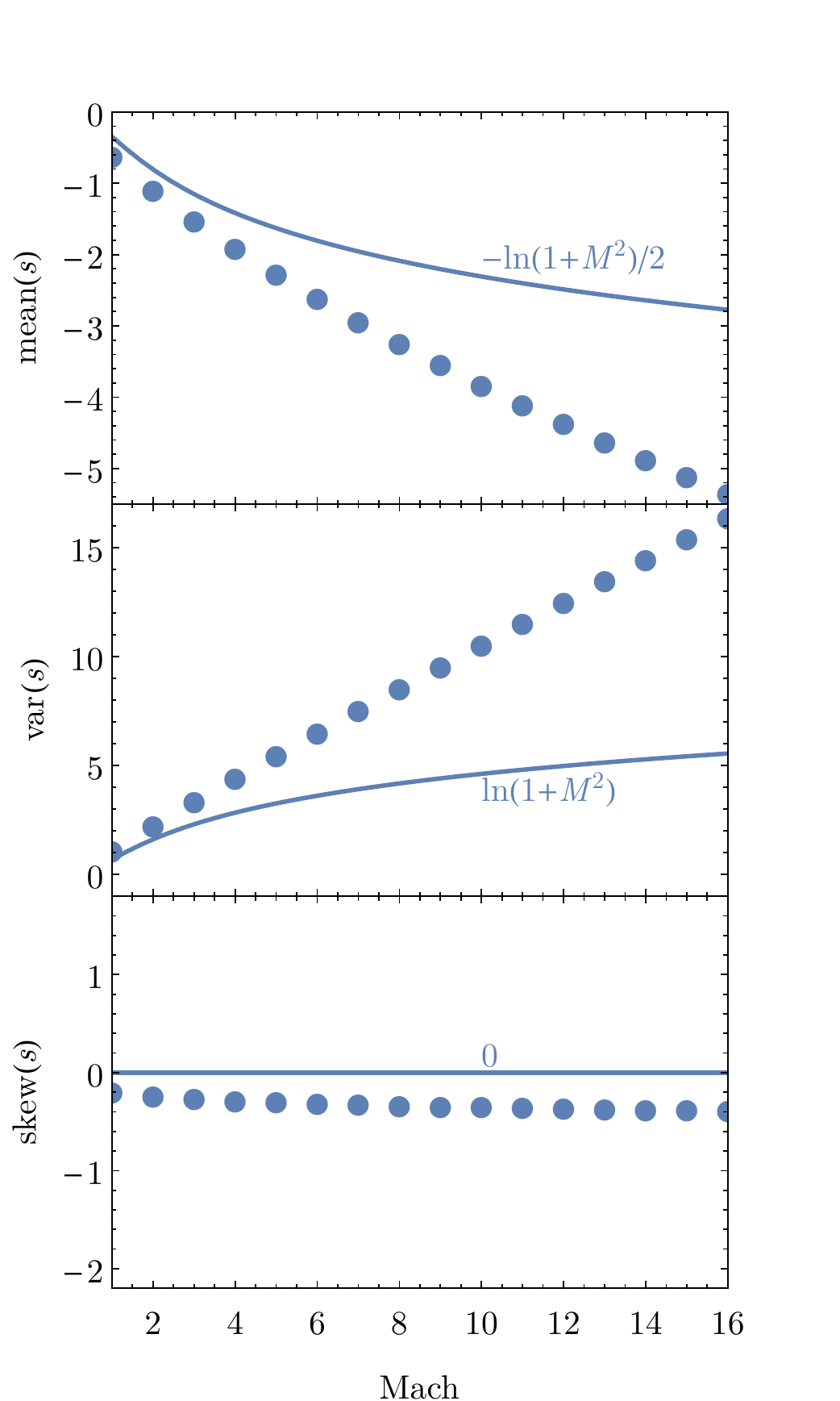}
\caption{Mean, variance, and skewness of the density PDF as a function of 1D (compressive) Mach number (dotted), and comparison to the lognormal model (thin).
With increased Mach number, the PDF becomes more negatively skewed as strong shocks make the turbulence more non-lognormal, and variance varies linearly with Mach number. \label{fig:stats}}
\end{figure}

\subsection{Non-lognormal PDF statistics}

Higher-order moments of the turbulent density distribution can reveal its underlying physical conditions and the nature of turbulent driving \citep{2009ApJ...693..250B}.
We calculate the resulting mean, variance, and skewness of our model as a function of 1D (compressive) Mach number, and compare it to the lognormal model, shown in Fig.~\ref{fig:stats}. Strong shocks cause non-lognormal density distributions, and evolve on faster timescales than the rest of the fluid, leading to a negative skewness that grows with Mach number.

The variance of the PDF as a function of Mach number is a highly discussed quantity in the study of turbulent statistics (see, e.g., the discussion around Fig.~6 of \citealt{2013MNRAS.430.1880H}). Across various turbulent simulations in the literature, it has been seen that the skewness as a function of Mach number does not vary simply as $\log(1+\mathcal{M}_{\rm compressive}^2)$, and it can in fact be larger, and there is large variance at high Mach numbers. The variance is a function of the degree of non-lognormality in the simulation, which can be affected by the details of the turbulent driving. The measured variance in simulations may also be affected due to limitations of resolution and finite box size (which bias the variance to slightly smaller values).
Interestingly, our Markov model predicts that when the turbulence is significantly non-lognormal, the variance scales more strongly than $\log(1+\mathcal{M}_{\rm compressive}^2)$ at high Mach numbers. In fact, it scales linearly with the Mach number (Fig.~\ref{fig:stats}). This linear scaling for variance in 1D simulations is a theoretical upper limit for 3D simulations, where rotational motions in turbulence may reduce the variance.

The mean of the PDF is simply set by mass conservation. Our non-lognormal Markov model has decreased mean log-density relative to the lognormal model.

\subsection{Comparison to other PDF models}

\cite{2018ApJ...865L..14S} have also recently considered a Markov model for explaining the PDF in turbulence, which serves as a general framework for our approach. There are a few important differences between our model and theirs.
We note that both models choose time scales that are phenomenological, based on simulation data. First, our model has an analytic solution, and describes the strong deviations from lognormal of the turbulence, while \cite{2018ApJ...865L..14S} choose $\tau_D(s)=\tau_A(s)$ purposefully to yield a near lognormal PDF, and their model has to be evaluated numerically through a Markov simulation. Second, we modeled the volume-averaged PDF with the timescales being measured from simulations in an Eulerian frame, while \cite{2018ApJ...865L..14S} modeled the mass-averaged PDF with timescales measured for Lagrangian volume elements, although the the PDFs can easily be related to each other. We note however that our timescales cannot directly be compared to \cite{2018ApJ...865L..14S}  because they are measured in different frames and  Eulerian quantities include the contribution of advection in changes in $s$, and hence will always measure larger changes over a fixed time interval. Third, we ensure exact mass conservation through proper renormalization of the PDF, whereas it is only approximate in \cite{2018ApJ...865L..14S}.

\cite{2013MNRAS.430.1880H,2017MNRAS.471.3753S} provide an analytic model for non-lognormal PDFs, motivated by a phenomenological scaling of structure functions in intermittent turbulence, characterized by a parameter $T$, where $T=0$ is lognormal, and $T=\infty$ is maximally intermittent.
It was found that the parameter $T$ increases systematically with the compressive Mach number
of turbulent simulations. As such, the PDF is described by a single parameter (the compressive Mach number), just as we have found. The model is complimentary to ours, where we have obtained the PDF by considering the timescales of transient structures, rather than their spatial distributions.
The $T$-models do have a time interpretation as well,
 discussed in \cite{2013MNRAS.430.1880H}: 
they represent the most general steady-state result 
 of a class of multiplicative-random-relaxation processes (with e.g. a 
 Poisson-distributed `waiting time' distribution between discrete 
 events that produce random multiplicative changes to the local value 
 then damp exponentially). 
 It is related to our model, except 
 for the assumptions about the discrete nature of `events' and damping.
 The $T$-model also has the nice mathematical feature that 
 when the density field is convolved on different scales with a filter, the
 field statistics of log-density remain exactly a $T$-model, as opposed to log-normal distributions where convolution can lead to small biases in the statistics.

\subsection{Applications of the PDF to star formation}

In the theoretical picture of a turbulent ISM, gas becomes gravitationally unstable beyond a critical density. Such a threshold has been investigated analytically and with simulations \citep{2005ApJ...630..250K,2011ApJ...743L..29H,2011ApJ...741L..22P,2012ApJ...761..156F,2017ApJ...838...40M,2018ApJ...863..118B,2019ApJ...879..129B}. 
The star formation efficiency per freefall time is the fraction of gas above this threshold.

For illustrative purposes, we take a transition density\footnote{Various treatments of the critical density in the literature may vary with order unity factors, which we ignore.
Alternatively one may forego the use of a transition density and compare timescales
$\tau(s)$ and $t_{\rm ff}(s)$ in a Lagrangian frame. But in our Eulerian frame $\tau(s)$ also includes advection which does not oppose gravitational collapse. So here we just use an  `illustrative' transition density, rather than one computed directly from the model.
} of $s_{\rm crit} = \log(\mathcal{M}^2)$. This is the post-shock density in the isothermal limit.  The post-shock density can be taken as a critical density for collapse in virialized cores without strong magnetic fields \citep{2005ApJ...630..250K,2011ApJ...741L..22P,2019ApJ...879..129B}.
The star formation efficiency (mass-weighted) per unit freefall time is estimated as:
\begin{equation}
    \epsilon = \int_{s_{\rm crit}}^{\infty}P_{\rm M}\,ds
    \label{eq:eps}
\end{equation}

In Eqn.~\ref{eq:eps}  there is no fudge-factor to account for the removal of gas by stellar winds or feedback.
We will examine here how non-lognormality affects star formation efficiency, which are observed to be a few percent \citep{2012ApJ...745...69K}.

Our definition of the SFE in Eqn.~\ref{eq:eps} is similar to that of \citet{2019ApJ...879..129B}, where the efficiency is estimated as the fraction of gas above the critical density, relative to the total gas.  The primary difference between our approach and that of \citet{2019ApJ...879..129B} is that the form of the density PDF is different. \citet{2019ApJ...879..129B} consider Eqn.~\ref{eq:eps} using a power law form of the density PDF, which is the expected form considering gas that is influence primarily by its own self-gravity. 
 In our model, there is no self-gravity and we only take into account the non-lognormality of the PDF.  
 We also do not include the `multi-free fall' density inside the integral as is done in studies such as \citep{2012ApJ...761..156F}, which give a very strong increase of the star formation efficiency per free fall with sonic Mach number.
 The `multi-free fall' scenario re-weights the integrand of Eqn.~\ref{eq:eps} by $t_{\rm ff}(\overline{\rho})/t_{\rm ff}(\rho)$.

\begin{figure}[ht!]
\plotone{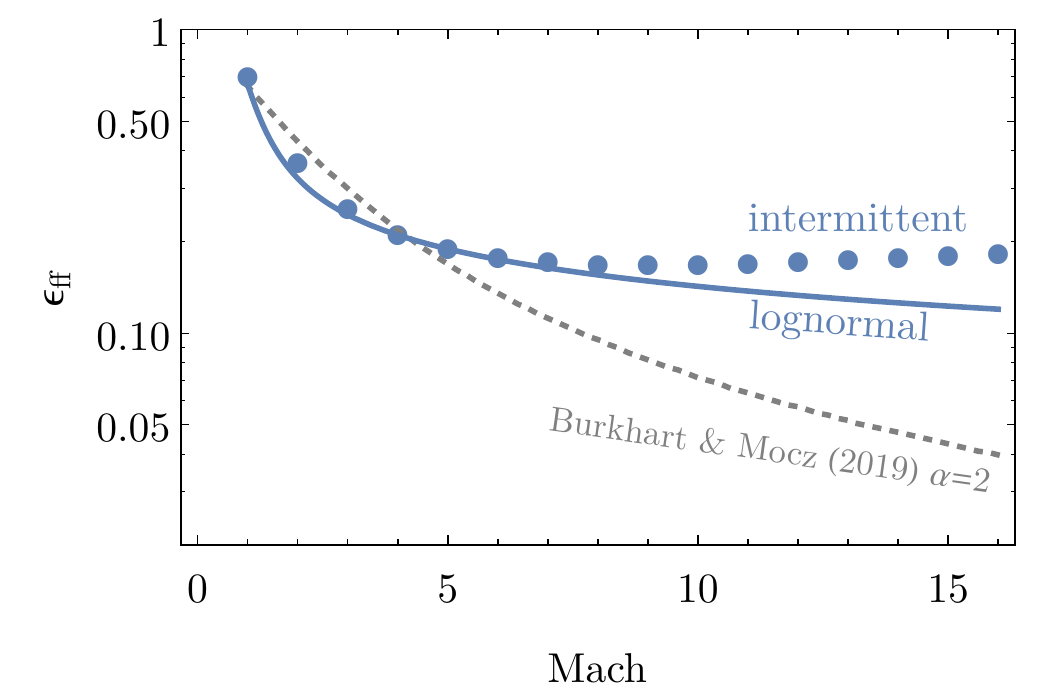}
\caption{Star formation efficiency (calculated as fraction of gas above critical density $s_{\rm crit}$) vs. sonic Mach number.   The efficiency is increased in our non-lognormal model for turbulence at high Mach number, due to the density PDF being negatively skewed. 
For comparison, we show efficiencies calculated from a lognormal model and the `lognormal plus ($\alpha=2$) powerlaw' model of \cite{2019ApJ...879..129B}.
The multi-free fall time factor \citep{2012ApJ...761..156F} is not included in this comparison, which can reverse the trend with Mach number.
\label{fig:ff}}
\end{figure}

Fig.~\ref{fig:ff} shows the star formation efficiency of our Markov model compared to the prediction for a lognormal, as a function of sonic Mach number. The star formation efficiency is increased in our model for turbulence at high Mach numbers, due to the negative skew of the PDF. If the degree of non-lognormality of star forming gas varies as a function of time, it means that the average star formation efficiency can change by a factor of $\sim 2$.  
The role of this non-lognormality may be incorporated into more detailed models of star formation efficiency.

\section{Conclusions}\label{sec:conc}

We have developed a simple Markov model with an analytic solution to describe the non-lognormal density distribution in supersonic, isothermal turbulence. The model was calibrated against 1D `turbulent' shock simulations, and shown to fit PDFs of 3D supersonic turbulence as well. Our main conclusions are as follows:
\begin{itemize}
    \item High density structures in supersonic turbulence are produced by strong shocks that have shorter transient timescales than low density structures.
    \item These strong shocks lead to deviations from Gaussianity in the natural log of the density PDF, particularly an increasing negative skewness as a function of turbulent Mach number.
    \item We have characterized these features of supersonic turbulence with a simple Markov model.
    \item Our results can be applied to both 1D (shocks only) and 3D simulations (shocks plus turbulence) with various driving methods (compressive vs solenoidal). This strongly suggests that the simplified 1D picture of shocks holds in 3D (which is a system of quasi-1D, interacting shocked sheet fragments with turbulence).
    \item Star formation efficiencies calculated from density PDFs are expected to increase at high Mach numbers when the PDF exhibits strong deviations from lognormal.
    \item Our model also applies to subsonic compressive turbulence, where the model reduces to a near lognormal density distribution.
\end{itemize}
Our model provides a simple way to quantify the non-lognormality of density structures in turbulent environments, such as star forming molecular clouds. It can be applied to factor in this feature in studies where the density PDF is used to predict star formation efficiencies.

\acknowledgments

The authors would like to thank Eve Ostriker, Jim Stone, and Mohammad Safarzadeh for valuable discussions.
Support for this work (P.M.) was provided by NASA through Einstein Postdoctoral Fellowship grant number PF7-180164 awarded by the Chandra X-ray Center, which is operated by the Smithsonian Astrophysical Observatory for NASA under contract NAS8-03060.
B.B. thanks the Simons Foundation for generous support.

\bibliography{mybib}

\end{document}